\documentclass[twocolumn,preprintnumbers,nofootinbib,prl]{revtex4}

\usepackage{graphicx}

\usepackage[hypertex]{hyperref}
\newcommand{\beq}{\begin{equation}}
\newcommand{\eeq}{\end{equation}}
\newcommand{\be}{B_\oplus}

\def\be{\begin{equation}}
\def\ee{\end{equation}}
\def\baray{\begin{eqnarray}}
\def\earay{\end{eqnarray}}
\def\ba{\begin{eqnarray}}
\def\ea{\end{eqnarray}}

\def\bk{{\bf k}}

\def\bx{{\bf x}}
\def\by{{\bf y}}

\def\CH{{\cal H}}
\def\CL{{\cal L}}

\begin{document}

\newcommand{\bea}{\begin{eqnarray}}
\newcommand{\eea}{\end{eqnarray}}
\newcommand{\barr}{\begin{array}}
\newcommand{\earr}{\end{array}}

\pagestyle{plain}

\preprint{MAD-TH-06-9}

\title{The Inflationary Trispectrum for Models with Large Non-Gaussianities}

\author{Xingang Chen$^1$, Min-xin Huang$^2$ and Gary Shiu$^2$}

\affiliation{
$^1$ Center for Theoretical Physics, Massachusetts Institute of
  Technology, Cambridge, MA 02139 \\
$^2$ Department of Physics, University of Wisconsin,
Madison, WI 53706, USA}


\begin{abstract}
We compute the leading order contribution to the four-point
function of the primordial curvature perturbation in a class of single field models where the inflationary
Lagrangian is a general function of the inflaton and its first
derivative. This class of models includes
 string motivated inflationary models such as
DBI inflation.
We find that the trispectrum for some range of parameters could potentially
be observed in future experiments. Moreover, the trispectrum can
distinguish DBI inflation from other inflation models with large non-Gaussianities
which typically have a similar bispectrum.
We also derive a set of
consistency conditions for $n$-point functions of the primordial
curvature perturbation in single field inflation,
generalizing Maldacena's
result for 3-point
functions.
\end{abstract}
\maketitle

The Cosmic Microwave Background (CMB)
provides us with a remarkably detailed snapshot of the early universe.
The vast structure that we see today is consistent at present with the simple picture of
an almost scale invariant,
Gaussian primordial density perturbation generated by inflation.
Precision measurements of
any small
deviation from
this simple picture
contain valuable information that
could constrain
cosmological models.

In the case of single field inflation, the non-Gaussian
perturbation is governed by the $n$-point functions ($n \geq 3$)
of the curvature perturbation $\zeta$. The leading non-Gaussian
features are known as the bispectrum (three-point function) and
trispectrum (four-point function), with their sizes conventionally
denoted as $f_{NL}$, $\tau_{NL}$ respectively. The current
experimental bound for the bispectrum from WMAP3 is
$-54<f_{NL}<114$ \cite{Spergel:2006hy}. Although the experimental
bound for the trispectrum is rather weak at the moment
$|\tau_{NL}|<10^8$ \cite{AL}, the next generation of experiments
such as PLANCK will increase the sensitivity to about
$|\tau_{NL}|\sim 560$ \cite{Kogo}, see also
\cite{Creminelli:2006}, \cite{Okamoto}. The CMB bispectrum and
trispectrum, if observed, could provide strong constraints on
inflationary models. In particular, the shape of the trispectrum
contains even more information about the inflationary dynamics
than the bispectrum because of the richer varieties of momentum
dependence it can encode. As we shall see, in some string
motivated inflationary models such as DBI inflation
\cite{Silverstein}, the size of the bispectrum and trispectrum can
both in principle reach the observable limit of future
experiments. Therefore it is crucial to study the shapes of
higher-point functions in such inflationary models.

The CMB bispectrum for single field inflation has been
studied intensively,
see, e.g.,
\cite{CHKS, Maldacena,Acquaviva:2002ud, Seery, Gruzinov}.
The
four-point function for slow roll inflation, however, has only recently been studied  \cite{Seery:2006},
where
it was found that the inflationary trispectrum generated by
slow roll models is too small to be observed even by future
experiments. In this note we compute the shape of the
trispectrum for DBI inflationary models and more generally a
class of single field models
considered in \cite{Garriga}
where the inflationary Lagrangian is a general function of the
inflaton and its first derivative. We find that for $n$-point functions,
there are $n-1$ independent shapes that could be potentially
large, providing useful information in distinguishing various
inflation models.
In the context of slow-roll inflation, Maldacena \cite{Maldacena} has
proposed an interesting consistency relation for the three point
function in a squeezed limit where one of the momentum modes has a
much longer wavelength than the other modes.
The importance of this relation is that its violation, if observed experimentally,
could rule out a large class of inflationary scenarios model-independently \cite{Creminelli:2004}.
In view of this, we derive a set of consistency
conditions for higher-point functions, generalizing Maldacena's result. Although some aspects of the
consistency conditions for the trispectrum in slow roll models have
recently been discussed in \cite{Seery:2006}, our
result is applicable to  a more general inflationary Lagrangian
and to cases when there are more
than one long wavelength modes. We show that the shapes of the trispectrum
we found satisfy these generalized consistency conditions. We end with
some final remarks and discuss the implications for future CMB
experiments.

Let us
begin with DBI inflation which was
proposed in \cite{Silverstein}
and studied subsequently in e.g. \cite{CHKS,Chen,Shiu,Shandera:2006ax}.
Although DBI inflation can be formulated without referring to branes
and extra dimensions, it is most easily visualized as describing a D-brane
moving relativistically in a warped throat where the inflaton $\phi$
is the position of the D-brane. For a warp factor $f(\phi)$
and an inflaton potential $V(\phi)$,
the inflationary Lagrangian is
\begin{eqnarray}
S && =\frac{M_{Pl}^2}{2}\int d^4x \sqrt{-g}R -\int d^4x
\sqrt{-g}\nonumber \\ &&
[f(\phi)^{-1}\sqrt{1-f(\phi)(\dot{\phi}^2-a(t)^{-2}(\nabla\phi)^2)}
\nonumber \\ && -f(\phi)^{-1}+V(\phi)] ~,
\end{eqnarray}
where $a(t)$ is the scale factor
and $H=\frac{\dot{a}(t)}{a(t)}$
is the Hubble parameter in the Friedmann-Robertson-Walker (FRW)
universe. The background solution for the inflaton
$\phi$ is spatially homogeneous. The sound speed, which describes the speed of fluctuations on this background, is defined as
\begin{eqnarray}
c_s \equiv \sqrt{1-f(\phi)\dot{\phi}^2} ~.
\end{eqnarray}
If the potential $V(\phi)$ is not flat, the inflaton will quickly
approach the relativistic speed limit $\dot{\phi}^2 \sim
f(\phi)^{-1}$ and so $c_s\ll 1$ . We will see that for DBI
inflation, the bispectrum and trispectrum scale with the sound
speed as:
\begin{eqnarray}
f_{NL}\sim \frac{1}{c_s^2},~~~~ \tau_{NL}\sim \frac{1}{c_s^4},
\end{eqnarray}
This can be easily understood as follows. By
expanding the DBI Lagrangian in the small sound speed limit, we
find that the $n$-point function scales like $\frac{1}{c_s^{3n-5}}$. Since
$f_{NL}$ is proportional to the three-point function divided by
the square of the power spectrum, it scales as
$\frac{1}{c_s^2}$. Similarly, $\tau_{NL}$ is proportional to
the four-point function divided by the cube of the power spectrum, and hence
it scales as $\frac{1}{c_s^4}$. Thus, there can exist models
with a sufficiently small sound speed $c_s$  that the size of the
trispectrum $\tau_{NL}$ is bigger than the sensitivity level of
future experiments $\tau_{NL}\geq 560$, but yet not too small to
be already ruled out by the WMAP bound on the CMB bispectrum
$f_{NL}$. As analogous to the bispectrum, we expect the bound on
$\tau_{NL}$ to depend on its shape \cite{Babich:2004gb} and so it
would be interesting to carry out a similar analysis for the
trispectrum. We note, however, that the bound on the trispectrum is
rather weak and so only in the small sound
speed limit and the leading contributions to $\tau_{NL}$ could possibly be observable
in future experiments.
In the small sound speed limit, the non-Gaussianities are dominated by
the inflaton's derivative interactions.
The contributions from the gravity sector and potential terms of the
scalar are suppressed by slow roll parameters, 
we will not
consider these subleading contributions in this paper. However, it is
of theoretical interests \cite{Seery:2006}
to study in more details these subleading
contributions since they may provide a testing ground
for theoretical issues such as the generalized
Maldacena's consistency conditions that we will present here.

We use the ``in-in'' formalism \cite{Weinberg:2005vy} 
to calculate the four-point expectation value for the primordial
curvature perturbation,
\bea
\langle \zeta^4 \rangle = -i \int_{t_0}^t dt'
\langle [\zeta^I(t,\bk_1) \cdots \zeta^I(t,\bk_4), H_{\rm int}(t')]
\rangle ~,
\label{4pt}
\eea
where $H_{\rm int}= \int d^3x \CH_{\rm int}$ is the interaction
Hamiltonian. Consider a perturbation $\alpha(t,\bx)= \delta \phi$
around the background solution.
We first work out the general relation between the interaction
Hamiltonian and the interaction Lagrangian up to the fourth order in
$\alpha$. We expand the fluctuation Lagrangian
density (starting from the second order)
to the fourth order in $\alpha(t,\bx)$, and denote it as
\bea
\CL &=& f_0 \dot \alpha^2 + j_2
+ g_0 \dot \alpha^3 + g_1 \dot \alpha^2 +
g_2 \dot \alpha + j_3 \nonumber \\
&+& h_0 \dot \alpha^4 + h_1 \dot \alpha^3 + h_2 \dot
\alpha^2 + h_3 \dot \alpha + j_4 ~,
\eea
where the $f_0$, $j$'s,
$g$'s and $h$'s are all functions of $\alpha(t,\bx)$, its
spatial derivative $\partial_i \alpha$ and $t$. The
subscripts 0, 1, 2, 3, 4 denote the orders of $\alpha$ in these
functions. 
We use the definition of the momentum density
\bea
\pi \equiv \frac{\partial
  \CL}{\partial \dot \alpha}
\eea
to express $\dot \alpha$ in terms of $\pi$ up to the third order,
\bea
\dot \alpha &=& \frac{\pi}{2f_0} + c_2 \pi^2 + c_3 \pi^3 ~,
\\
c_2 &=& - \frac{3g_0}{8f_0^3} - \frac{g_1}{2 f_0^2 \pi} -
\frac{g_2}{2f_0\pi^2} ~, 
\\
c_3 &=& -\frac{1}{2f_0} ( \frac{3g_0c_2}{f_0} + \frac{h_0}{2f_0^3})
\nonumber \\
&-& \frac{1}{2f_0 \pi} (2g_1c_2 
+ \frac{3h_1}{4f_0^2})
- \frac{h_2}{2f_0^2 \pi^2} - \frac{h_3}{2 f_0 \pi^3} ~,
\eea
and plug it into the Hamiltonian density
\bea
\CH = \pi \dot \alpha - \CL ~.
\eea
We then separate the $\CH$ into a kinematic Hamiltonian density 
$\CH_0$ which
is quadratic in $\pi$ and $\alpha$, and an interaction Hamiltonian
density $\CH_{\rm int}$. 
To use the ``in-in'' formalism, the $\alpha$ and $\pi$ in this
interaction Hamiltonian should be replaced, respectively, 
by the $\alpha^I$ and
$\pi^I$ in the interaction picture. These latter variables satisfy the free
equation of motion (in the time-dependent background) followed from 
$\CH_0$ and the usual
commutation relation
\bea
[\alpha^I(t,\bx), \pi^I(t,\by)]= i\delta^3(\bx-\by) ~.
\label{commutation}
\eea
Expressing the $\pi^I$ in $\CH_{\rm int}$ in terms of $\dot
\alpha^I$ using
\bea
\dot \alpha^I = \frac{\partial \CH_0}{\partial \pi^I} 
= \frac{\pi^{I}}{2f_0} ~,
\eea
we finally get (omitting the label ``$I$'' in the variables in
$\CH_{\rm int}$ from now on)
\bea
\CH^{\rm int}_{3} &=&  - g_0 \dot \alpha^3 - g_1 \dot \alpha^2 -g_2
\dot \alpha   - j_3  ~, \\
\CH^{\rm int}_4 &=& ( \frac{9 g_0^2}{4 f_0} - h_0) \dot
\alpha^4 
+ (\frac{3g_0 g_1}{f_0} - h_1)  \dot \alpha^3
\nonumber \\
&+& ( \frac{3g_0 g_2}{2 f_0} + \frac{g_1^2}{f_0} - h_2)
\dot \alpha^2
+( \frac{g_1 g_2}{f_0} - h_3) \dot \alpha 
\nonumber \\
&+& \frac{g_2^2}{4f_0} - j_4 ~.
\label{H4L4}
\eea
As we see, while the cubic part of $\CH_{\rm int}$ is the negative of
that of $\CL_{\rm int}$, this is generally not so for higher
order expansion. 
For slow-roll inflation, the extra terms are suppressed by slow-roll
parameters \cite{Adshead:2008gk}, and therefore can be properly neglected in
the trispectrum calculation of Ref.~\cite{Seery:2006}.
As we will see, for the case of interest here, the
extra terms contribute to the leading order results.

For the special case of the DBI action,
to compute the leading
shape of the four-point function, 
we only keep the leading terms in the small sound speed
limit, which come from expanding the square root term in the DBI
action. The Lagrangian density up to quartic order is:
\begin{eqnarray}
\mathcal{L}_2 & =&
\frac{a^3}{2c_s^3}[\dot{\alpha}^2 -a^{-2}c_s^2(\nabla\alpha)^2] ~,
\label{quadraticterm}
\\
\mathcal{L}_3 &=&
\frac{a^3}{2c_s^5\dot{\phi}}[\dot{\alpha}^3-a^{-2}c_s^2\dot{\alpha}(\nabla\alpha)^2] ~,
\\ \label{quarticterm}
\mathcal{L}_4 &=&
\frac{a^3}{8c_s^7\dot{\phi}^2}[5\dot{\alpha}^4-6a^{-2}c_s^2\dot{\alpha}^2(\nabla\alpha)^2\nonumber
\\ && +a^{-4}c_s^4(\nabla\alpha)^2(\nabla\alpha)^2] ~.
\end{eqnarray}
Using (\ref{H4L4}) we then have
\bea
\CH_4^{\rm int} = \frac{a^3}{2 c_s^7 \dot \phi^2} \dot\alpha^4 ~.
\label{Hint4}
\eea

 From the quadratic term we can solve for the perturbation and
quantize it according to the standard procedures of quantum field
theory:
\begin{eqnarray}
\alpha(t,\textbf{x})&=&\frac{1}{(2\pi)^3}\int d^3k
[u(\tau,{\textbf{k}})a(\textbf{k}) \nonumber \\ &&
+u^*(\tau,{-\textbf{k}})a^{\dagger}(-\textbf{k})]e^{i\textbf{k}\cdot\textbf{x}}~,
\end{eqnarray}
where
$u(\tau,{\textbf{k}})=\frac{H}{\sqrt{2k^3}}(1+ikc_s\tau)e^{-ikc_s\tau}$
is the solution of the quadratic Lagrangian, and
$\tau=-\frac{1}{aH}$ is the conformal time. The creation and
annihilation operators satisfy the usual commutation relation
$[a(\textbf{k}),a^{\dagger}(\textbf{k}^{\prime})]=(2\pi)^3\delta^3(\textbf{k}-\textbf{k}^{\prime})$,
consistent with (\ref{commutation}).

In the small sound speed limit, the primordial density
perturbation $\zeta$ is dominated by the fluctuation of the inflaton
$\alpha$, and is simply related to it by
$\zeta=\frac{H}{\dot{\phi}}\alpha$ \cite{Silverstein}. The two
point function is
\begin{eqnarray}
\langle\zeta(\textbf{k}_1)\zeta(\textbf{k}_2)\rangle=(2\pi)^5\delta^3(\textbf{k}_1+\textbf{k}_2)\frac{P^\zeta_{k_1}}{2k_1^3},
\label{2pt}
\end{eqnarray} where $P^\zeta_k=\frac{H^4}{(2\pi)^2 \dot{\phi}^2}$ is
the power spectrum, and we have evaluated the perturbation at several
e-foldings after the horizon exit with the approximation
$\tau\rightarrow 0$.

Using (\ref{Hint4}), the four-point function at leading
order in $\zeta$ is
\bea
&&\langle \zeta(\bk_1) \zeta(\bk_2) \zeta(\bk_3) \zeta(\bk_4) \rangle
\nonumber \\
&=& (2\pi)^9 \delta^3 (\sum_i \bk_i) (P^\zeta_K)^3 \frac{1}{c_s^4}
\frac{1}{\prod_i k_i^3} (\frac{1}{2} A_1 ) ~,
\label{3ptDBI}
\eea
where
\bea
A_1 &=& -72 \frac{\prod_i k_i^2}{K^5} ~.
\label{shape1} 
\eea
Here $P^\zeta_K$ is the power
spectrum at the leading order in slow variation parameters. 
Momentum conservation implies that the momenta $k_i$  where
$i=1 \dots 4$
form a quadrilateral. The quadrilateral may be skew, i.e. not lying on a plane. For convenience we will just refer to it as a quadrilateral.
This is in contrast to the bispectrum where the momenta form a triangle.

In order to compare with the value of $\tau_{NL}$ from
experiments, we need to specify a configuration of the
quadrilateral to evaluate the four-point function. We consider the
``equilateral configuration'' in \cite{Seery:2006}, where the $4$
momenta have the same amplitudes $k$, and hence the angles between them
satisfy $\cos(\theta_{12})=\cos(\theta_{34})$,
$\cos(\theta_{13})=\cos(\theta_{24})$, $\cos(\theta_{14})
=\cos(\theta_{23})$, and $\sum_{m=1}^3 \cos(\theta_{m4}) = -1$.
Here $\theta_{ij}$ is the angle between
$\textbf{k}_i$ and $\textbf{k}_j$.
Using the definition of $\tau_{NL}$
\begin{eqnarray}
&& \langle \zeta(\bk_1) \zeta(\bk_2) \zeta(\bk_3) \zeta(\bk_4) \rangle
\nonumber \\
&=& \frac{\tau_{NL}}{2}(2\pi)^9 \delta^3 (\sum_i \bk_i) \nonumber \\
&\times & [\tilde P_\zeta(k_1) \tilde P_\zeta(k_2) \tilde
  P_\zeta(k_{14})+23 ~\textrm{permutations} ] ~,
\label{taudef}
\end{eqnarray} 
where $\tilde P_\zeta(k_1)=P^{\zeta}_{k_1}/(2k_1^3)$ as defined in
(\ref{2pt}) and $\bk_{ij}=\bk_i+\bk_j$, we find that for DBI
inflation, by matching (\ref{3ptDBI}) and (\ref{taudef}) in the special
``equilateral configuration'',
\begin{eqnarray}
\tau_{NL} &=& -\frac{9}{128}\frac{2^{\frac{3}{2}}}{c_s^4
  \sum_{m=1}^3(1+\cos\theta_{m4})^{-\frac{3}{2}}} ~.
\end{eqnarray}
For a generic angular configuration, we can roughly estimate that
$\tau_{NL}\sim \frac{0.04}{c_s^4}$. This confirms our claim that
there are models that give rise to observable trispectrum, but are
not ruled out by current bounds on the bispectrum,
e.g., when
$c_s\sim 0.05$.

It is straightforward to generalize the analysis to a class of
models where the inflation Lagrangian is a general
function $P(X,\phi)$, with
$X=-\frac{1}{2}g^{\mu\nu}\partial_{\mu}\phi\partial_{\nu}\phi$
\cite{Garriga} . The
sound speed is defined as
\begin{eqnarray}
c_s^2=\frac{P_{,X}}{P_{,X}+2XP_{,XX}}~.
\end{eqnarray}
Again we consider only the leading contribution in the small sound
speed limit. In this case the derivative of the Lagrangian with $\phi$,
i.e., $P_{,\phi}$, does not give rise to shapes large enough to be observed. One can
show that the leading order expansions up to quartic order are
\begin{eqnarray}
\CL_2 &=& \frac{1}{2} ( \frac{\partial P}{\partial X} + \frac{\partial^2
  P}{\partial X^2} \dot \phi^2) a^3 \dot \alpha^2 
-\frac{1}{2} \frac{\partial P}{\partial X} a (\nabla \alpha)^2 ~,
\label{generalL2} \\
\CL_3 &=& (\frac{1}{2} \frac{\partial^2 P}{\partial X^2} \dot \phi +
\frac{1}{6} \frac{\partial^3 P}{\partial X^3} \dot \phi^3) a^3 \dot
\alpha^3  \nonumber \\
&& -\frac{1}{2} \frac{\partial^2 P}{\partial X^2} \dot\phi~ a \dot
\alpha (\nabla \alpha)^2 ~, 
\label{generalL3} \\
\mathcal{L}_4 &=&
a^3[\frac{1}{24}\dot{\phi}^4\frac{\partial^4P}{\partial
X^4}\dot{\alpha}^4+\frac{1}{4}\dot{\phi}^2\frac{\partial^3P}{\partial
X^3}\dot{\alpha}^2(\dot{\alpha}^2-a^{-2}(\nabla\alpha)^2)\nonumber
\\ && +\frac{1}{8}\frac{\partial^2P}{\partial
X^2}(\dot{\alpha}^2-a^{-2}(\nabla\alpha)^2)^2] ~.
\label{generalL4}
\end{eqnarray}
So for the four-point function 
we found 3 independent shapes (i.e., momentum dependences). A convenient  basis
is the
shape $A_1$ we
found earlier, and two additional shapes coming from terms
proportional to $a \dot \alpha^2 (\nabla \alpha)^2$ and $a^{-1}
(\nabla \alpha)^2 (\nabla \alpha)^2$, respectively, in
$\CH^{\rm int}_4$,
\bea
A_2 &=& - \frac{1}{4} \frac{k_1^2 k_2^2 (\bk_3\cdot \bk_4)}{K^3}
\left( 1+ \frac{3(k_3+k_4)}{K} + \frac{12 k_3 k_4}{K^2} \right) 
\nonumber \\
&+& {\rm perm.} ~,
\label{shape2} \\
A_3&=&
-\frac{1}{4}\frac{(\textbf{k}_1\cdot\textbf{k}_2)(\textbf{k}_3\cdot\textbf{k}_4)}{K}(1+\frac{\sum_{i<j}k_ik_j}{K^2}
\nonumber \\ && +\frac{3k_1k_2k_3k_4}{K^3} (\sum_i\frac{1}{k_i})
+12\frac{k_1k_2k_3k_4}{K^4}) \nonumber \\ && +\textrm{perm.} ~,
\label{shape3}
\eea
where $K=\sum_{i=1}^4 k_i$ and ``perm.'' refers to the other 
$23$ permutations
of the four momenta.
The coefficients of these shapes are determined by
the Lagrangian (\ref{generalL2})-(\ref{generalL4}) 
and the relation (\ref{H4L4}).

The shapes of the four-point function $A_1, A_2, A_3$ can be used to
distinguish between models with different parameters
$\frac{\partial^4P}{\partial X^4}$, $\frac{\partial^3P}{\partial
X^3}$, $\frac{\partial^2P}{\partial X^2}$. To illustrate this, we can
compare the situation
with that of the bispectrum, where the shapes were computed in
\cite{Gruzinov, CHKS}. There it was found that at the leading
order there are two large shapes of non-Gaussianities, denoted by
$A_{\lambda}$ and $A_c$:
\begin{eqnarray}
A_{\lambda}&\sim& \frac{k_1^2k_2^2k_3^2}{K^3}, \nonumber \\
A_c & \sim &
-\frac{1}{K}\sum_{i>j}k_i^2k_j^2+\frac{1}{2K^2}\sum_{i\neq
j}k_i^2k_j^3
+\frac{1}{8}\sum_ik_i^3.
\end{eqnarray}
However, these two shapes are functions of the momentum triangle and look
qualitatively similar in plots \cite{CHKS}. In particular, in the
squeezed limit $k_1\rightarrow 0$, one can easily show these two
shapes have the same asymptotic behaviors $A_{\lambda}, A_c \sim
k_1^2$. The shapes of the four-point function Eqs.~(\ref{shape1}),(\ref{shape2}),(\ref{shape3})
are more distinguishable. For
example, we can take a squeezed limit $k_1\rightarrow 0$, and the
remaining three sides of the quadrilateral form a triangle. The
asymptotic behaviors of the shapes are
\begin{eqnarray}
A_1 &\sim& k_1^2(\frac{k_2^2k_3^2k_4^2}{K^5}) ~, \nonumber \\
A_2 &\sim&
\frac{k_1k_2k_3k_4}{K^3}[k_3k_4\cos(\theta_{12})(1+\frac{3k_2}{K})
\nonumber \\ && +k_2k_4\cos(\theta_{13})(1+\frac{3k_3}{K})
\nonumber \\ && +k_2k_3\cos(\theta_{14})(1+\frac{3k_4}{K})] ~,
\nonumber \\
A_3 &\sim&
\frac{k_1k_2k_3k_4}{K}(1+\frac{k_2k_2+k_2k_4+k_3k_4}{K^2}+\frac{3k_2k_3k_4}{K^3})
\nonumber \\ &&
[\cos(\theta_{12})\cos(\theta_{34})+\cos(\theta_{13})\cos(\theta_{24})
\nonumber \\ && +\cos(\theta_{14})\cos(\theta_{23})] ~.
\end{eqnarray}
In the squeezed limit of a generic configuration of the
quadrilateral,  the first shape scales as $A_1\sim k_1^2$, and two
other shapes scale as $A_2, A_3\sim k_1$. So this in principle
provides a
more refined
discriminator between
different inflation models
than the bispectrum.
Another
interesting configuration is
when
$\textbf{k}_1$,
$\textbf{k}_2$ and $\textbf{k}_3$
are
orthogonal to each other and
$\textbf{k}_4=-\textbf{k}_1-\textbf{k}_2-\textbf{k}_3$. In this
configuration, the shape $A_3$
 vanishes. In
particular, these provide the distinction between DBI inflation
and the power law k-inflation model in
\cite{Armendariz-Picon:1999rj}. The simplest power law k-inflation
has a Lagrangian $P(X,\phi) \sim (-X + X^2)/\phi^2$.
It was shown in
\cite{Garriga, Armendariz-Picon:1999rj} that the model has a power
law inflationary solution and the inflaton moves at constant
speed. Using the interaction Hamiltonian (\ref{H4L4}), we find the fourth order Hamiltonian has three terms proportional to 
$a^3\frac{\partial^2 P}{\partial X^2} \dot{\alpha}^4$, $a\frac{\partial^2 P}{\partial X^2} \dot{\alpha}^2(\nabla\alpha)^2$,
$a^{-1}\frac{\partial^2 P}{\partial X^2}
(\nabla\alpha)^2(\nabla\alpha)^2$. We know for example from the
calculations in the DBI case that the $\dot{\alpha}^4$,
$a^{-2}c_s^2\dot{\alpha}^2(\nabla\alpha)^2$,
$a^{-4}c_s^4(\nabla\alpha)^2(\nabla\alpha)^2$ give rise to the same
$c_s$ order contribution to trispectrum in the small sound speed
$c_s$ limit. So we can see the $A_3$ shape coming from the interaction
term $(\nabla\alpha)^2(\nabla\alpha)^2$ dominate over the other shapes
for the simple power law k-inflation. On the other hand, the DBI 
inflation does not have the $A_2$ and $A_3$ component and their features
in the above momentum configurations. 
So these distinguish the power law
k-inflation from the DBI inflation.

One can go to more complicated corners of the configuration space
of the quadrilateral and study the different behaviors of the
shapes $A_1$, $A_2$, $A_3$, but it should be clear that the
configuration space of a quadrilateral is much bigger than that of
a triangle, and thus provides a more powerful discriminator for
inflationary models.

Now we derive a generalized consistency condition for $n$-point
function, and test it in various models for which we have computed
the four-point function. In single field inflation, the density
perturbation is generated by a scalar perturbation $\zeta$
whose $l$-point function is:
{\small
\begin{eqnarray}
\langle \zeta(\textbf{k}_1)\cdots\zeta
(\textbf{k}_l)\rangle=(2\pi)^3\delta^3(\textbf{k}_1+\cdots+\textbf{k}_l)P^{l}(\textbf{k}_1,\cdots,\textbf{k}_l)
\end{eqnarray}}
The scalar perturbation $\zeta$ has scaling behavior of degree
$-3$, so the $l$-point function
$P^{l}(\textbf{k}_1,\cdots,\textbf{k}_l)$ is roughly a degree
$3-3l$ homogeneous function of $\textbf{k}_i$, $i=1,\cdots,l$. We
can define the spectral index of $l$-point functions analogously
to that of the two-point function
\begin{eqnarray}
n_l-1&=&\frac{d\textrm{log}(P^{l}(\textbf{k}_1,\cdots,\textbf{k}_l))}{d\textrm{log}(k)}+3l-3
\nonumber \\ &=&
\frac{d\textrm{log}(\langle\zeta(\textbf{k}_1)\cdots\zeta(\textbf{k}_l)\rangle)}{d\textrm{log}(k)}+3l
\end{eqnarray}
In the above definition all the $\textbf{k}_i$ are taken to be of the same order
of magnitude as $k$, e.g. $\frac{d \log (k_i)}{d \log (k)}=1$.
When $l=2$, this is just the usual spectral index of the power
spectrum $n_s-1$. Another example is $l=3$.
In this case the three-point spectral index is
\begin{eqnarray}
n_3-1=\frac{d\textrm{log}(f_{NL})}{d \textrm{log}(k)}+2(n_s-1)
\end{eqnarray}
where $\frac{d\log (f_{NL})}{d \log (k)}$ is known as the running
of non-Gaussianity \cite{Chen:2005fe}.  In an inflationary model where the physical
parameters vary slowly within a Hubble time, the $l$-point
spectral index is very small, i.e. $|n_l-1|\ll 1$.

We will take the squeezed limit of the $n$-point function,
where $l$ of the
momenta ($2\leq l<n$) are much bigger than the others, i.e.,
\begin{eqnarray}
k_1,\cdots,k_l\gg k_{l+1},\cdots,k_n~.
\end{eqnarray}
The long wavelength modes $k_{l+1},\cdots,k_n$ exit the horizon
and
are
frozen much earlier than the short wave modes
$k_{1},\cdots,k_l$. Their effect is to act as a background
that scales the spatial coordinates, since they perturb the FRW
metric as $ds^2=dt^2-a(t)^2e^{2\zeta_B} d \textbf{x}^2$. The $l$-point
function in a background perturbation $\zeta_B\sim 10^{-5}$ is
\begin{eqnarray}
\langle\zeta(\textbf{x}_1)\cdots\zeta(\textbf{x}_l)\rangle_B=
\langle\zeta(e^{\zeta_B}\textbf{x}_1)\cdots\zeta(e^{\zeta_B}\textbf{x}_l)\rangle~.
\end{eqnarray}
The background $\zeta_B$ is roughly constant around the short wave
length scale $|x_i-x_j|$ with $1\leq i<j\leq l$, and can be
evaluated at
$\textbf{x}_0=\frac{\textbf{x}_1+\cdots+\textbf{x}_l}{l}$. We
expand the above equation to first order in the background
\begin{eqnarray}
&&  \langle\zeta(\textbf{x}_1)\cdots\zeta(\textbf{x}_l)\rangle_B =
\langle\zeta(\textbf{x}_1)\cdots\zeta(\textbf{x}_l)\rangle
\nonumber \\ &&+\zeta_B(\textbf{x}_0)\{(
\frac{d}{d\zeta_B}\langle\zeta(e^{\zeta_B}\textbf{x}_1)
\cdots\zeta(e^{\zeta_B}\textbf{x}_l)\rangle)|_{\zeta_B=0}\}
\end{eqnarray}
The zeroth order term is independent of the background so will not
contribute to the final result. After a Fourier transformation to
momentum space, the first order term in the background expansion
becomes
\begin{eqnarray}
&&\langle\zeta(\textbf{k}_1)\cdots\zeta(\textbf{k}_l)\rangle_B
\nonumber \\ 
&=& \int \frac{d^3\textbf{k}_B}{(2\pi)^3} \zeta_B (\textbf{k}_B)
\{\frac{d}{d\zeta_B}(e^{-3l\zeta_B}
\langle\zeta(e^{-\zeta_B}(\textbf{k}_1-\frac{\textbf{k}_B}{l}))
\nonumber \\ &&
\cdots\zeta(e^{-\zeta_B}(\textbf{k}_l-\frac{\textbf{k}_B}{l}))\rangle)|_{\zeta_B=0}\}
\end{eqnarray}
Here we introduce an integration variable $\textbf{k}_B$, which will be fixed by the delta function to be
 $\textbf{k}_B \equiv \textbf{k}_1+\cdots + \textbf{k}_l$, and which 
satisfies
$k_B\ll k_i, i=1,\cdots
l$. At the leading order of the squeezed limit we can neglect the
background momentum in the $l$-point function, and use our
definition of the $l$-point spectral index to find
\begin{eqnarray}
\langle\zeta(\textbf{k}_1)\cdots\zeta(\textbf{k}_l)\rangle_B &=&
-(n_l-1) P^l(\textbf{k}_1,\cdots,\textbf{k}_l)  \nonumber
\\ & &\times \zeta(\textbf{k}_1+\cdots+\textbf{k}_l)
\end{eqnarray}
Plugging in the expression for the $n$-point function, we find a
generalization of Maldacena's consistency condition to $n$-point
function in the squeezed limit $k_1,\cdots,k_l\gg k_{l+1},\cdots,k_n$:
\begin{eqnarray} \label{npointconsistency}
&& P^n(\textbf{k}_1,\cdots,\textbf{k}_n) = - (n_l-1)P^l(\textbf{k}_1,\cdots,\textbf{k}_l)\nonumber \\
&& \times
P^{n-l+1}(\textbf{k}_1+\cdots\textbf{k}_l,\textbf{k}_{l+1},\cdots,\textbf{k}_n)
\end{eqnarray}

In the case of DBI inflation, the $n$-point function on the left
hand side of the consistency condition (\ref{npointconsistency})
scales as $\frac{(P^\zeta_k)^{n-1}}{c_s^{2n-4}}$, while the right
hand side scales as $\frac{(P^\zeta_k)^{n-1}}{c_s^{2n-6}}$, so the
$n$-point function must vanish in the squeezed limits at leading
order for small sound speed.  We see indeed the shapes $A_1$,
$A_2$ and $A_3$ in (\ref{shape1}), (\ref{shape2}), (\ref{shape3})
all vanish in a squeezed limit when there is a long wave mode
$k_i\rightarrow 0$, in agreement with the $n$-point consistency
condition (\ref{npointconsistency}). A more non-trivial check of
(\ref{npointconsistency}) is when
$P^n(\textbf{k}_1,\cdots,\textbf{k}_n)$ is non-vanishing in the
squeezed limit, e.g., one may check the validity of
(\ref{npointconsistency}) using the recent results for slow roll
inflation \cite{Seery:2006} (together with subleading terms which
are yet to be determined). Besides serving as a check on the
calculations, the generalized consistency condition
(\ref{npointconsistency}) if violated can, in the spirit of \cite{Creminelli:2004},
rule out a large class of models experimentally.

For a general Lagrangian $P(X,\phi)$,
the $n$-point function
has $n-1$ independent shapes proportional to the
derivatives of the Lagrangian $\frac{\partial ^iP}{\partial X^i}$,
where $i=2,3, \cdots, n$. It is not difficult to compute the shapes
and show that they vanish in the squeezed limit, though it is difficult to measure such
higher point functions experimentally.

We emphasize that the CMB trispectrum, although difficult to observe, could provide a useful
discriminator for inflationary models. In particular,
if large non-Gaussianities are observed experimentally in the future, it may still be difficult to
distinguish via the bispectrum DBI inflation from other inflationary models governed by derivative interactions, e.g., the models in \cite{Gruzinov}.
The trispectrum we computed may serve as an extra filter to zero-in on string inflationary scenarios.

We thank
D.~Chung, G.~Geshnizjani,
S.~Kachru,
B.~Underwood, P.~Adshead, F.~Arroja, R.~Easther, K.~Koyama, E.~Lim,
and D.~Seery
for discussions. We thank Kazuya Koyama and Shuntaro Mizuno for
spotting an error in a numerical factor in the v3 of the paper.
This work was supported in part by NSF CAREER Award No. PHY-0348093,
DOE grant DE-FG-02-95ER40896, a Research Innovation
Award and a Cottrell Scholar Award from Research Corporation. XC is
supported by the US Department of Energy under cooperative research
agreement DEFG02-05ER41360.

\end{document}